\documentclass{article}
\usepackage{ijcai18}

\usepackage{graphicx}
\usepackage{url}
\usepackage{times}
\usepackage{xcolor}
\usepackage{soul}
\usepackage[utf8]{inputenc}
\usepackage[small]{caption}
\usepackage{nicefrac}
\usepackage{cleveref}
\usepackage{float}

\usepackage{caption}
\usepackage{subcaption}

\title{Balanced News Using Constrained Bandit-based Personalization}

\author{
Sayash Kapoor$^{\star\dagger}$, 
Vijay Keswani$^\star$, 
Nisheeth K. Vishnoi$^\star$,
L. Elisa Celis$^\star$
\\ 
$^\star$ {\'Ecole Polytechnique F\'ed\'erale de Lausanne (EPFL), Switzerland}\\
$^\dagger$ Indian Institute of Technology Kanpur, India\\
}

\begin{document}

\maketitle
\begin{abstract}
We present a prototype for a news search engine that presents \emph{balanced} viewpoints across liberal and conservative articles with the goal of de-polarizing content and allowing users to escape their filter bubble. 
The balancing is done according to flexible user-defined constraints, and leverages recent advances in constrained bandit optimization.
We showcase our balanced news feed by displaying it side-by-side with the news feed produced by a traditional (polarized) feed.

\end{abstract}

\section{Introduction}

The personalization of content online can lead to opinion bubbles and polarization \cite{FGR16}.
As a user clicks on  content of a particular type (e.g., links from liberal media outlets), the content delivery system learns these preferences, and in the future presents the user with more items of that type. 
In doing so, many content delivery engines polarize completely,
displaying a very narrow range of content to each user.
This can lead to polarization and the formation of opinion bubbles
in which people are only exposed to content that matches their pre-existing beliefs \cite{Matakos2017}, 
and can lead people to mistrust information that does not match their opinion \cite{RibeiroCAM17}.
Personalization leading to polarization has been observed in social media feeds \cite{FGR16}, product recommendation \cite{MGGR15}, online advertising \cite{DTD2015} and other media that pervades the internet.
Moreover, diverse search results might be desirable from the point-of-view of user satisfaction \cite{CelisTekriwal2017}.
Our goal in this work is to show that show that an alternative exists: Balanced content delivery systems are feasible, and one can even give the user the information and freedom to control their balance of content. 
In this demo, as an example, we take news search in the US diversified across liberal and conservative content.

\begin{figure}[!t]
    \centering
    {\includegraphics[width=0.4\textwidth]{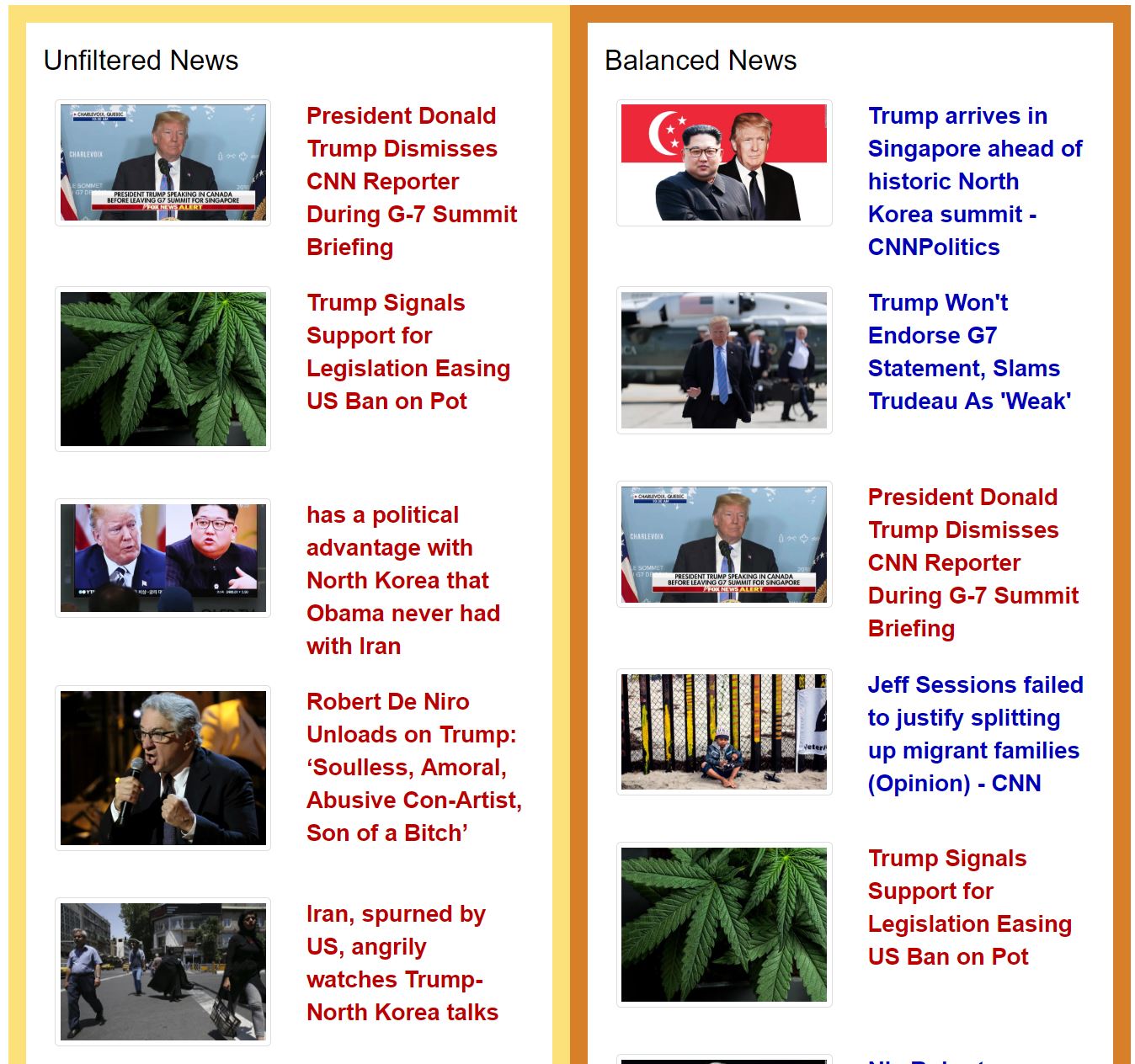}}
    \caption{{Balanced news search demo for a liberal-favoring user.} A traditional system ({unfiltered news}) on the left would learn about the user and only display liberal-favoring articles. Our constrained system ({balanced news}) on the right similarly learns the user preferences, but still displays some conservative articles in order to provide more diversified content. \emph{(Liberal-favoring articles in blue, conservative-favoring articles in red.)}
    }
    \label{fig:screenshots}

\end{figure}

\textbf{Balanced News} is our prototype
\footnote{Corresponding author: L. Elisa Celis.}
\footnote{\emph{Demo video}: {http://y2u.be/Zvum0t1EYtQ}} \footnote{\label{demowebsite}\emph{Demo website}: {https://bit.ly/2H1vroP}}
 aimed at demonstrating novel bandit-based \emph{constrained} personalization algorithms
that can be used for delivering diverse content while still allowing for personalization.
A feature of this recent work is that the algorithms are scalable despite constraints, allowing for real-time systems to be balanced in this manner.
The algorithms optimize content in the presence of constraints,
that is, limits on how much or how little content of a particular \emph{type}  is displayed. 
We consider news search in the US and classify the content into two \emph{types}:
liberal-favoring and conservative-favoring.
\begin{figure}[t]
\centering
    {\includegraphics[width=\linewidth]{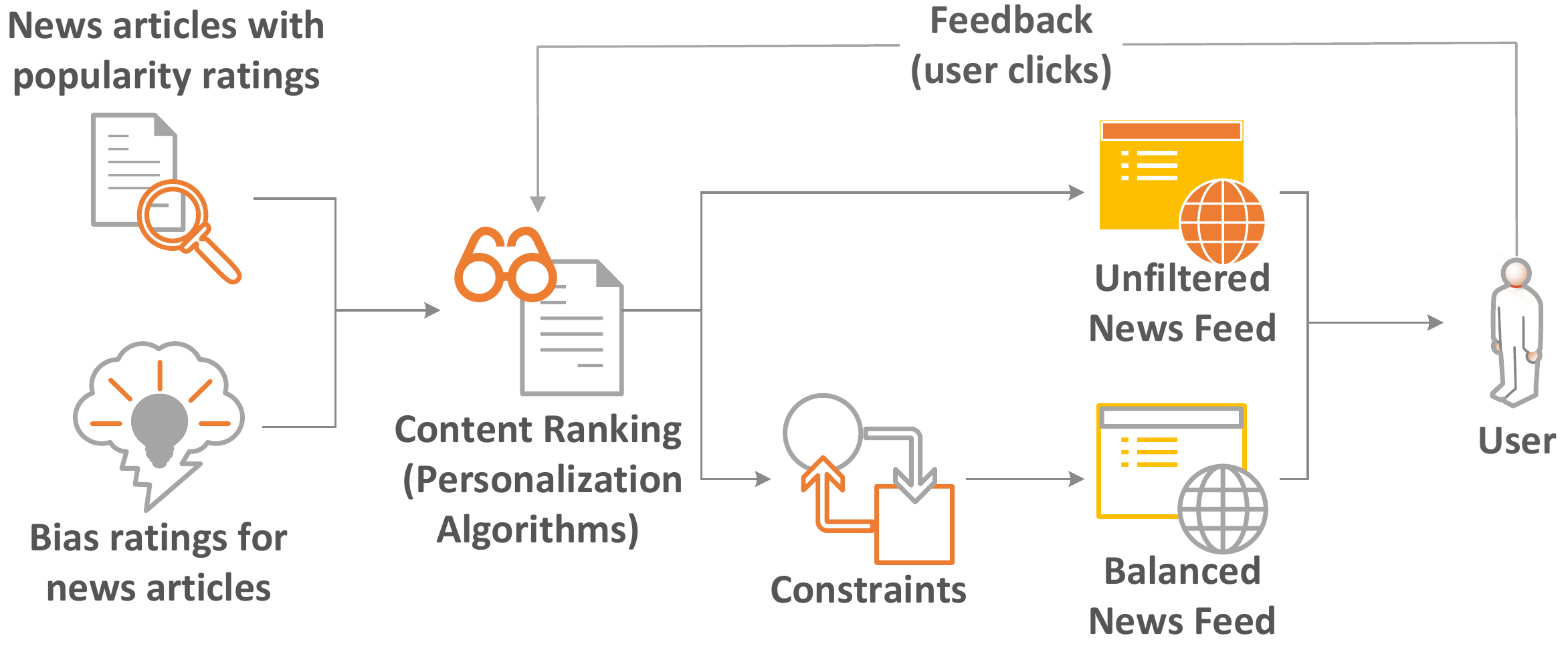}}
    \caption{
{Flowchart of the balanced search prototype.} }
    \label{fig:flow-chart}
\end{figure}
\Cref{fig:screenshots} shows a screenshot of the demo website.
For showcasing the types that each article belongs to, 
articles belonging to the {liberal} type have blue hyperlinks,
whereas articles belonging to the {conservative} type have red hyperlinks.
There are two news feeds shown in the demo -- the \emph{unfiltered} feed that runs a standard Bandit algorithm used for personalization,
and the \emph{balanced} feed that runs the {constrained} Bandit algorithm from \cite{CV17,CKSV18}.
In both feeds, clicking on conservative articles decreases the proportion of liberal articles in the next iteration.
Similarly, clicking on liberal articles increases the proportion of liberal articles.
The main difference between the unfiltered and balanced news feeds is that
while the unfiltered feed can continue getting polarized in a direction untill \emph{all} articles belong to only one type,
the balanced news feed can only specialize up to user-defined constraints -- 
once the constraints have been reached, the proportion of articles of the type favored by the user can not increase.

On a high-level, similar ``group-fairness'' constraints have also been considered for other fundamental problems, including classification~\cite{dwork2012fairness,zafar2017fairness,CHKV18},  data summarization~\cite{celis2016fair,CKSDKV17}, ranking~\cite{celis2018ranking,yang2017measuring}, and voting \cite{celis2017group}.

\section{Implementation}
In our demo we display two different news feeds side-by-side in order to contrast the difference between a traditional (unfiltered) news feed, and our (balanced) news feed. The former is implemented using standard bandit algorithms, and the latter
 is implemented using {constrained} bandit algorithms as inspired by ~\cite{CKSV18}.
For implementing the algorithms, for each article we require article a \emph{rating}, a \emph{type} (either liberal or conservative) and a \emph{reward} function.
For the balanced algorithm, we also require \emph{constraints} on each type of content.

\noindent\textbf{Ratings.} We used
the \emph{WebHose News Search API},
\footnote{{https://webhose.io}}
which ranks articles by popularity.
This gives a warm-start to the Bandit algorithm; 
in the beginning we display articles based on popularity, and gradually personalize to the user's preferences as they are learned.

\noindent\textbf{Type.} For each article, the \emph{type} -- either liberal or conservative -- is determined based on bias ratings from \emph{AllSides}.
\footnote{{https://bit.ly/2GX1lRx}}

\noindent\textbf{Reward.} We take every user click as an indicator of a user's preference for viewing more articles of that type,
and use this as the \emph{reward} for the bandit algorithm.
	An article $a$ that the user clicks at the $t$-th iteration receives a reward $r_t^a =1$, 
	all other articles $a^\prime$ receive a reward $r_t^{a^\prime} = 0$.

\begin{figure}[t]
    \centering
    \fbox{\includegraphics[width=0.45\textwidth]{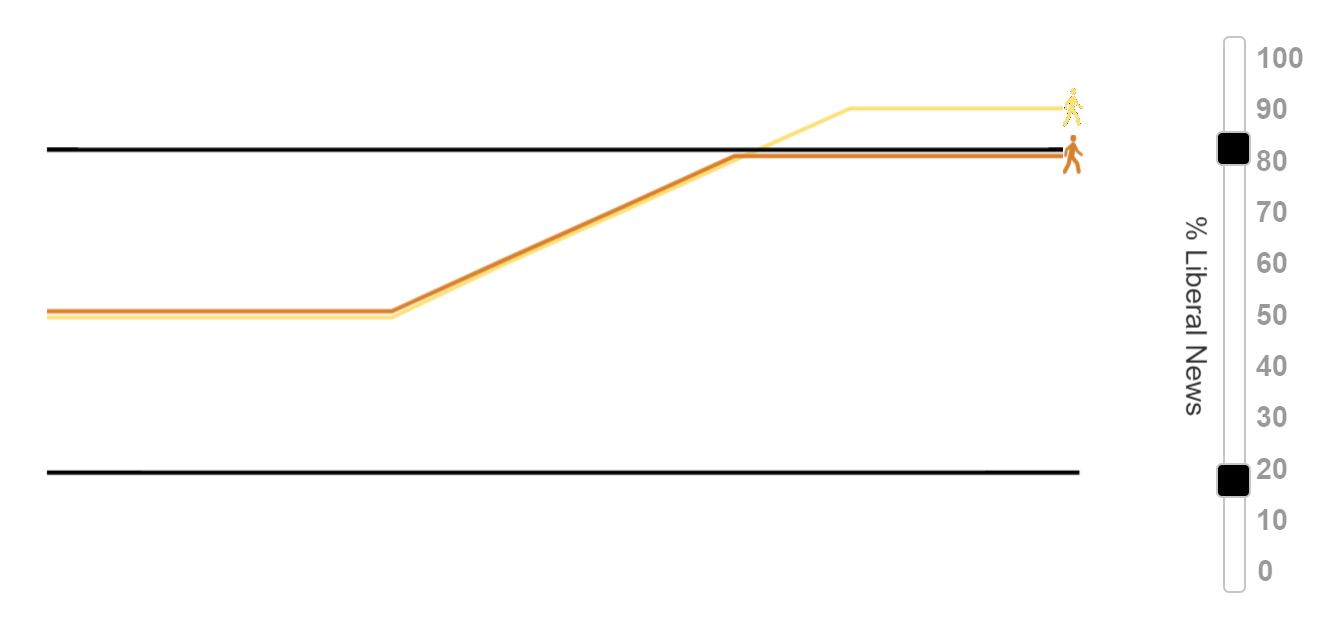}}
    \caption{{Dashboard displaying the percentage of liberal articles in the unfiltered and balanced news feeds.} Unfiltered news in yellow, balanced news in orange, constraints in black. While the percentage of liberal articles can increase (or decrease) to 100\% (0\%) in the case of unfiltered news, they are constrained to lie within thegiven region for balanced news \emph{(orange)}. The sliders on the right allow the user to change the constraints. 
    }
    \label{fig:news-search-constraints}
\end{figure}

\noindent\textbf{Constraints.} We take upper and lower constraints on the percentage of liberal articles as input through the slider to the right of the graph (see~\Cref{fig:news-search-constraints}). 
	By default, the constraints are set to $80\%$ and $20\%$ respectively; i.e., the balanced news feed would always contain \emph{at least} 20\% liberal articles, and 20\% conservative articles -- the remaining 60\% would depend on the given user's preference history.

The difference between the unfiltered and balanced news feeds becomes apparent once a user has clicked on enough articles of just one type:
While unfiltered news completely polarizes to display articles largely of a single type (either conservative or liberal, depending on user preferences),
balanced news always has a certain minimum percentage of articles of both types (as defined by the constraints) (see~\Cref{fig:screenshots}).

\Cref{fig:news-search-constraints} demonstrates an instance of the constraints in action by showing the outcome for a new user who, upon entering the system, clicks on 5 consecutive liberal articles, and no conservative articles.
The yellow line depicts the percentage of liberal articles in the unfiltered news feed over time,
the orange line depicts the percentage of liberal articles in the balanced news feed over time, and
the black lines depict the constraints over time.
For a user who has clicked a large number of liberal-favoring articles, 
the unfiltered news feed has high chances of displaying liberal-favoring articles,
whereas this probability is constrained by a certain upper (and lower) bound for the balanced news.

\section{Conclusion}
Using constrained personalization algorithms allows us to increase diversity and reduce polarization in content delivery systems,
which can help combat the formation of opinion bubbles.
Our balanced search prototype \footnotemark[3] is aimed at showcasing what such content delivery systems might look like. 
An important open problem is to determine how to set constraints on articles being delivered to users as soliciting constraints from the user directly may not always be feasible or desireable.
In particular, it would be important to determine which constraints provide enough diversity to prevent opinion bubbles from forming or content from having disparate impact on specific populations, while at the same time not compromising user satisfaction.
\newpage

\bibliographystyle{named}
\bibliography{references}
\end{document}